\documentclass[dvips]{article}
\usepackage{amsmath,amssymb}
\usepackage{icrctc07}
\title{Monte Carlo Study of Cosmic-Ray Propagation in the Galaxy and Diffuse Gamma-Ray Production}
\shorttitle{Monte Carlo Study of Cosmic-Ray Propagation}
\authors{C.-Y. Huang and M. Pohl}
\shortauthors{C.-Y. Huang and M. Pohl}
\afiliations{Department of Physics and Astronomy, Iowa State University, Ames, IA 50011}
\email{huangc@iastate.edu}

\abstract{This work presents preliminary results for the time-dependent cosmic-ray propagation in the Galaxy by a fully 3-dimensional Monte Carlo simulation.
The distribution of cosmic-rays (both protons and helium nuclei) in the Galaxy is studied on various spatial scales for both constant and variable
cosmic-ray sources. The continuous diffuse gamma-ray emission produced by cosmic-rays during the propagation is evaluated.}

\begin{document}
\maketitle
%Begin the section.

\subsection{Introduction}
One of the important goals of current cosmic-ray studies is to understand the origin as well as the propagation of cosmic-rays, in particular the sources of cosmic-ray 
particles and also the distribution of sources in the Galaxy. One widely used method is to examine the matter path 
length, or age distribution of cosmic-rays \cite{Biermann01,Sciama00}. Originally, the most popular model to describe the cosmic-ray propagation in the Galaxy was 
the leaky-box model (LBM) \cite{Engelmann90} that unrealistically assumes a uniform source distribution and the diffusion and boundary effect are 
taken into account by means of simple loss terms. Diffusion models \cite{Berezinskii90} are more realistic but their analytic forms are limited in the description of 
the many parameters required and their spatial and energy dependence that are needed to fully describe cosmic-ray propagation in the Galaxy. A numerical solution of the
diffusion equation by a finite-difference scheme avoids this problem, but is only accurate if the grid is much finer than all physical scales \cite{Strong98}. 
A Monte Carlo approach should be ideal for the study of cosmic-ray propagation in the vicinity of their sources. 

This work presents preliminary results for the time-dependent cosmic-ray propagation in the Galaxy by a fully 3-dimensional Monte Carlo simulation.
The distribution of cosmic-rays in the Galaxy is studied on various spatial scales for both constant and variable
cosmic-ray sources. The continuous diffuse gamma-ray emission during the cosmic-ray propagation is evaluated. 
\subsection{Model and Simulation Technique}
In the simulation, the cosmic-ray diffusion is described by a Monte Carlo process in which the particle position at time $t_{n+1}$ is related to its position at $t_n$ by
\begin{eqnarray}
x_i(t_{n+1}) =x_i(t_n) + \Delta x_i =x_i(t_n) +\cos\alpha_i \overline{\Delta r}
\end{eqnarray}
where $x_i$ as the particle 3-coordinate, $\Delta x_i$ the displacement for each coordinate, $\cos\alpha_i$ the direction cosine for each direction determined by a 
random number generators and $\overline{\Delta r}$ the mean displacement that is described as
\begin{eqnarray}
\overline{\Delta r} =\sqrt{6D(E)\Delta t}
\end{eqnarray}
with $D(E)$ as the energy-dependent diffusion coefficient and the timestep $\Delta t$. The choice of $\Delta t$ directly determines the spatial resolution.

The simulation methodology for the particle motion through physical space and energy space is as follows. After reading the physical configurations such as 
Galactic dimension, ISM distribution, Galactic magnetic field, total propagation time, detection shell dimension, source dimension and injection profile 
(energy range, spectral profile, etc.), a particle, after been injected by a source, is allowed to undergo a fully 3-dimensional random walk. At each step, the particle 
experiences energy loss and interactions. The particle continues to propagate until either it reaches the boundary, or the propagation time scale is greater than 
$10^8~\textrm{years}$, or its energy is below the minimum simulation energy (i.e., $E_{\textrm{min}}=m_p+0.3~\textrm{GeV}\simeq 1.24~\textrm{GeV}$). 
After reaching the end of the simulated propagation of a particle, the next particle is simulated. This technique allows the parallel computation without communication.      

\subsection{Result Examples from Simulation}
The transport equation of cosmic rays is generally written as 
\begin{eqnarray}\label{EQ:Diffuse}
& &\frac{\partial N}{\partial t}
=  Q(E,\mbox{\boldmath$r$},t) + \nabla \cdot (D\nabla N_i) -p_iN_i\nonumber\\
&+&  \frac{\partial}{\partial E} \left[b_i(E)N_i(E)\right] - \nabla \cdot (VN_i(E))  \nonumber\\
&+& \frac{v\rho}{m}\sum_{k\ge i}\int\frac{d\sigma_{i,k}(E,E')}{dE}N_{k}(E')dE'
\end{eqnarray}
where $N_i(E,\mbox{\boldmath$r$},t)dE$ is the particle number density of type $i$ at position $\mbox{\boldmath$r$}$ with energy in $(E,E+dE)$. Other terms, e.g. for
stochastic reacceleration, can be added as desired. For a simple test of the simulation code, we simplify Eq.~(\ref{EQ:Diffuse}) by keeping only the source term and the 
diffusion term:
\begin{eqnarray}\label{EQ:DiffuseSimple}
\frac{\partial N}{\partial t} 
=  Q(E,\mbox{\boldmath$r$},t) + \nabla \cdot (D\nabla N). 
\end{eqnarray}
Without losing generality, the solution to Eq.~(\ref{EQ:DiffuseSimple}) can be written as 
\begin{eqnarray}\label{EQ:DiffuseSimpleSol}
N (E,\mbox{\boldmath$r$},t) 
= \int\!\!\!\int\!\!\!\int dV'\,dE'\,dt'\, Q\cdot G
% = \int\!\!\!\int\!\!\!\int dV\,dE'\,dt'\, Q(E',\mbox{\boldmath$r$}',t') G(E',\mbox{\boldmath$r$'},t'). 
\end{eqnarray}
by the Green's function such that 
\begin{eqnarray}\label{EQ:GreenFnSimple}
G(E,\mbox{\boldmath$r$},t) =\frac{1}{8(\pi D t)^{3/2}} \exp\left[ -\mbox{\boldmath$r$}^2/(4Dt)\right]
\end{eqnarray}
where $G(\mbox{\boldmath$r$},t)$ gives the probability of finding a particle at position $\mbox{\boldmath$r$}$ and at time $t$ that was injected at $\mbox{\boldmath$r$'}=0$
and $t'=0$.
\begin{figure}[t]
\begin{center}
\vspace*{-1.2cm}
\includegraphics[width=6.8cm]{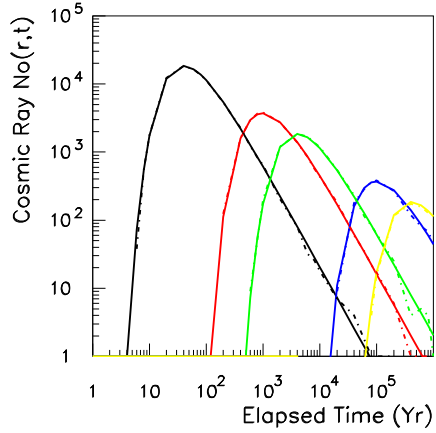}
\end{center}
\vspace*{-0.8cm}
\caption{Analytic solutions (solid lines) and simulation results (dashed lines) for the diffusion equation (\ref{EQ:DiffuseSimple}). We count particles in shells of width 1
pc, that are located at $|\mbox{\boldmath$r$}|=10,~50,~100,~500~\textrm{and}~1000~\textrm{pc}$. A total of $10^5$ particles are simulated.} 
\label{Fig:DiffuseSimple}
\end{figure}

Fig.~\ref{Fig:DiffuseSimple} shows the simulation results (dashed lines) and the analytic solutions 
(solid lines) calculated by Eqs.~(\ref{EQ:DiffuseSimpleSol}) and (\ref{EQ:GreenFnSimple}) for the particle number in different shells, following the instantaneous injection
of $10^5$ particles at the origin.

We also consider a case with continuous energy loss by synchrotron radiation and inverse Compton scattering, and an energy-dependent diffusion coefficient $D(E)=D_0E^{\delta}$
for the case of Kolmogorov-type spectrum of turbulence \cite{Heinbach95}. Thus the transport equation can be reduced to 
\begin{eqnarray}\label{EQ:DiffuseESynchrotron}
\frac{\partial N}{\partial t} - \frac{\partial}{\partial E} (b_0E^2N)-D_0E^{\delta}\nabla^2N =Q
\end{eqnarray}
with the source term 
\begin{eqnarray}\label{EQ:CRSource}
Q=Q_0 E^{-\alpha}\Theta(\tau-t_0)\Theta(t_0+T-\tau). 
\end{eqnarray}
The analytic solution  to Eq.~(\ref{EQ:DiffuseESynchrotron}) is obtained as \cite{Pohl98}
\begin{eqnarray}\label{EQ:SolESynchrotron}
N  = Q_0 E^{-\alpha} \int_{-1/b_0E}^0 d\tau \quad\quad\quad\quad\quad\quad\quad\quad\quad\nonumber \\
\cdot \left[\frac{\Theta(\tau-t_0) \Theta(t_0+T-\tau)\exp(-\mbox{\boldmath$r$}^2/4\Lambda)}{(4\pi\Lambda)^{3/2}(1+b_0E\tau)^{2-\alpha}}\right]
\label{9}
\end{eqnarray}
where the cooling time $\tau$ and the length scale $\Lambda$ are 
\begin{eqnarray}\label{EQ:CoolingTime}
\tau = \int_E^{E_0} \frac{dE'}{b_0E'^2}\, , \quad \Lambda = \int^{E_0}_E \frac{D(E')dE'}{b_0E'^2}.
\label{10}
\end{eqnarray}

\begin{figure}[t]
\begin{center}
\vspace*{-1.2cm}
\includegraphics[width=6.8cm]{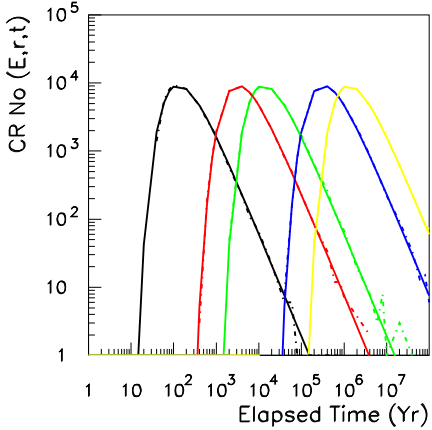}
\end{center}
\vspace*{-0.8cm}
\caption{Analytic solutions (solid lines) and simulation results (dashed lines) for the diffusion equation 
(\ref{EQ:DiffuseESynchrotron}) for $10^5$ particles with a same 
energy injected by the source. The figure is arranged as Fig.~\ref{Fig:DiffuseSimple} but with shell width 
of 10\% of $|\mbox{\boldmath$r$}|$ for each shell for better statistics.} 
\label{Fig:DiffuseESynSimple}
\end{figure}

Fig.~\ref{Fig:DiffuseESynSimple} shows the simulation results (dashed lines) and the analytic solutions (solid lines) to 
Eq.~(\ref{EQ:DiffuseESynchrotron}) with an injection of total $10^5$ particles with an initial energy $E=200~\textrm{GeV}$ by a source at $t=0$ and 
at the origin. The figure is arranged as Fig.~\ref{Fig:DiffuseSimple}.

\begin{figure*}[t]
\vspace*{-1.20cm}
\begin{center}
\hspace*{0.0cm}\mbox{\includegraphics[width=6.8cm]{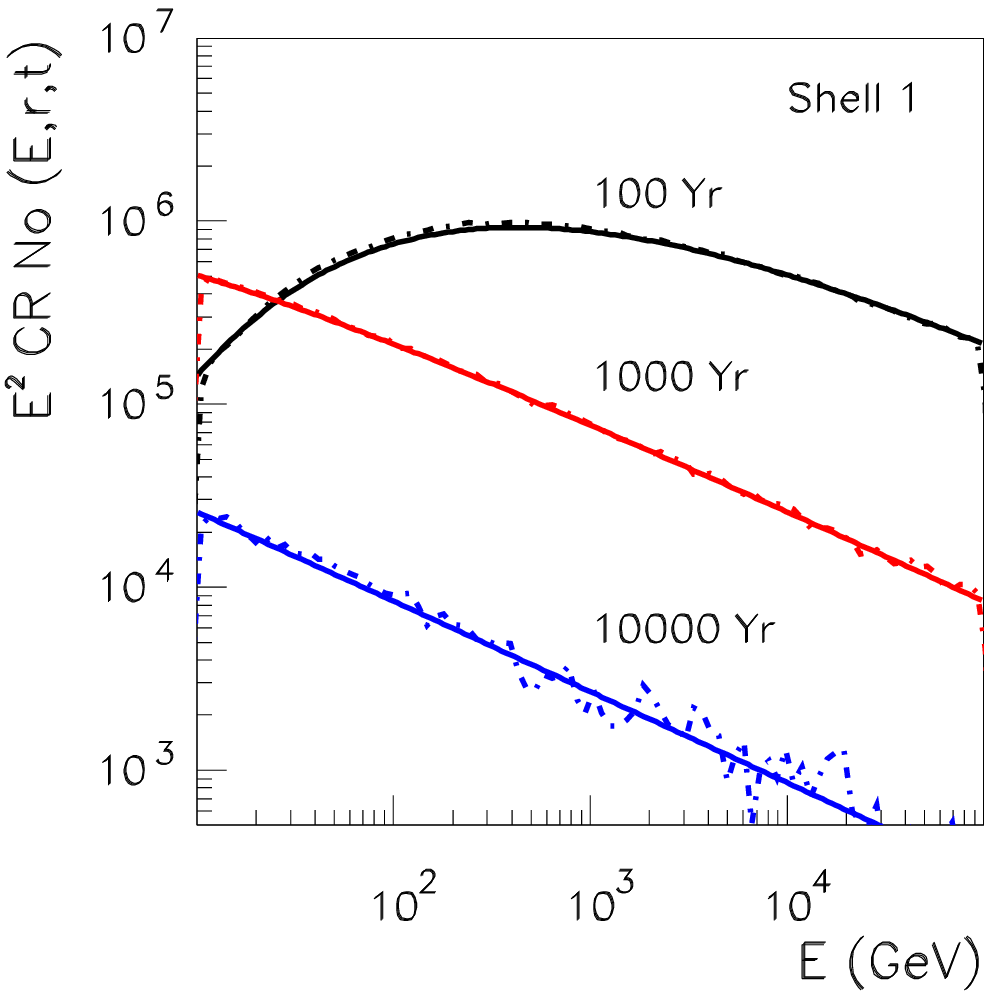}
\hspace*{0.0cm}      \includegraphics[width=6.8cm]{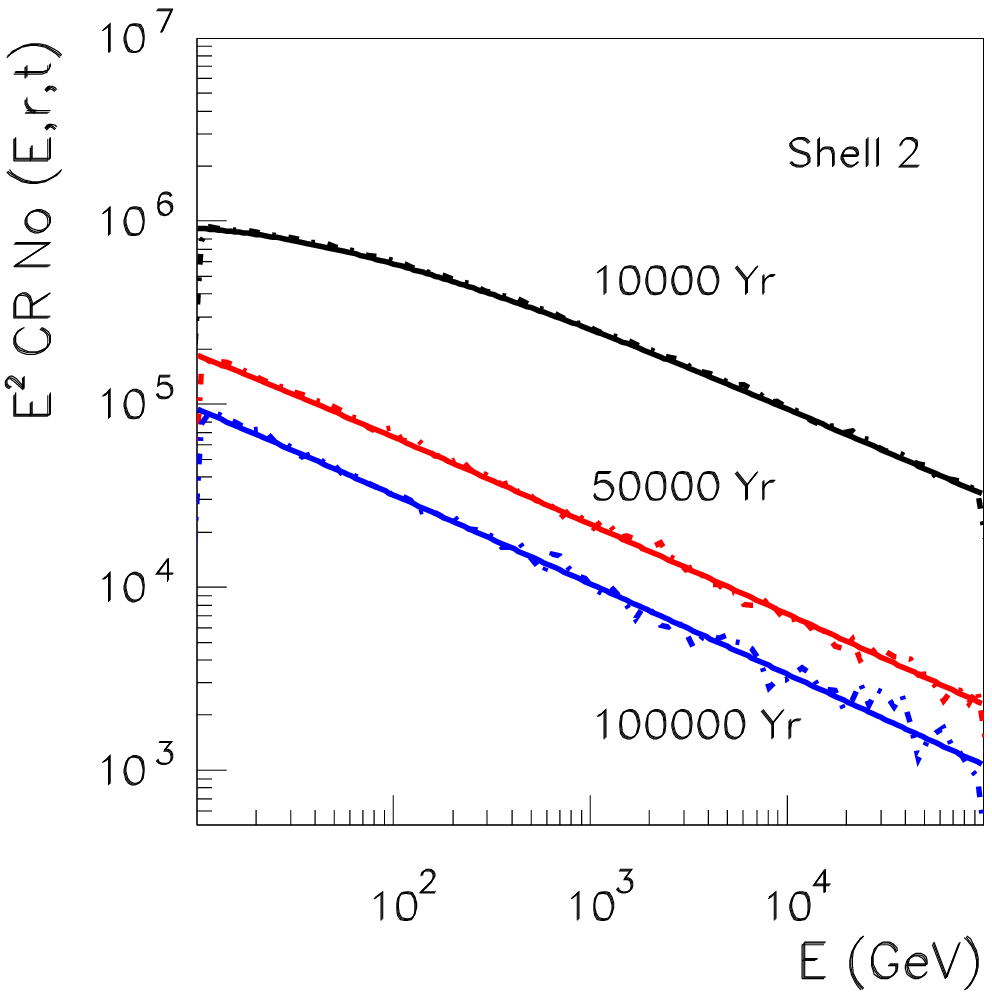}}
\vspace*{-0.4cm}
\caption{Particle distribution calculated by simulation (dashed lines) and analytic solutions (solid lines) for a power-law injection by a source 
at the origin and at $t=0$. Figure shows examples for 2 shells as defined in Fig.~\ref{Fig:DiffuseSimple} but with shell widths 20\% of $|\mbox{\boldmath$r$}|$ 
for different elapsed time: in range of $10^2$-$10^4$ years (shell 1) and $10^4$-$10^5$ years (shell 2).}\label{Fig:DiffusePowerLaw}
\vspace*{-0.8cm}
\hspace*{0.0cm}\mbox{\includegraphics[width=6.8cm]{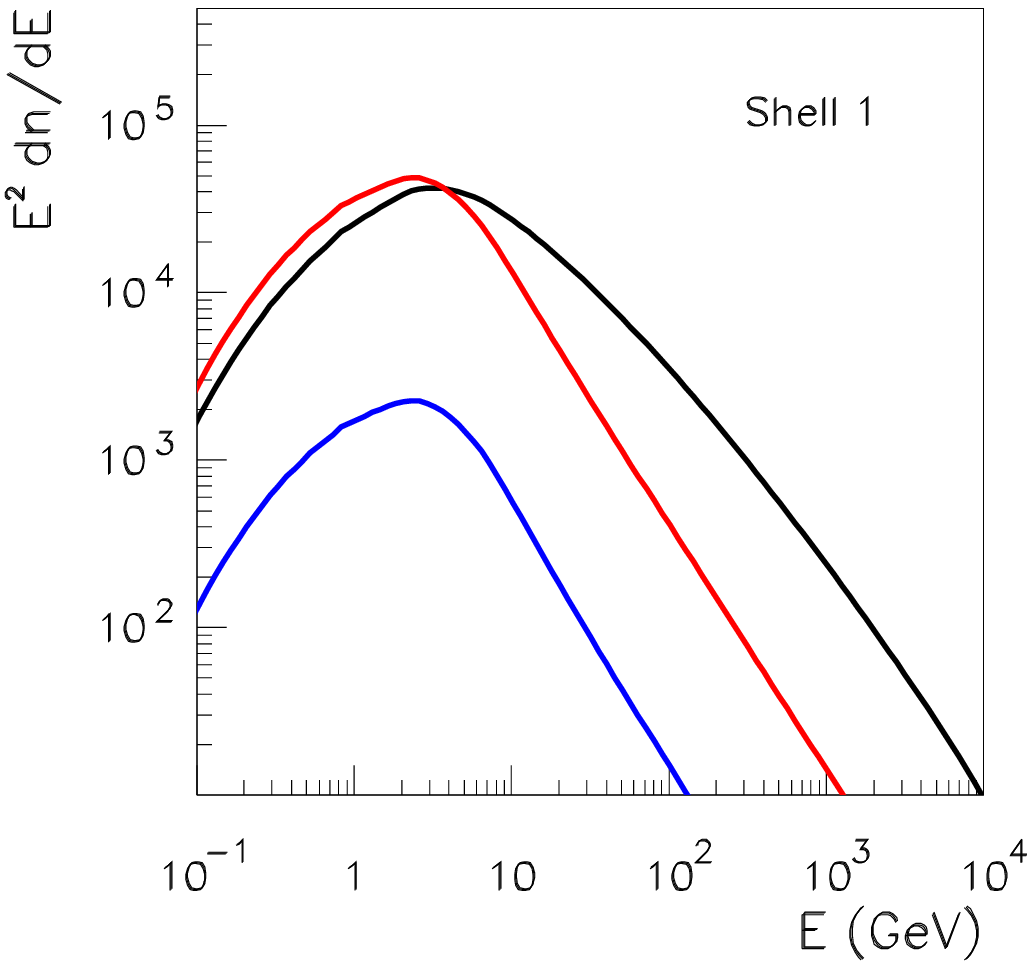}
\hspace*{0.0cm}      \includegraphics[width=6.8cm]{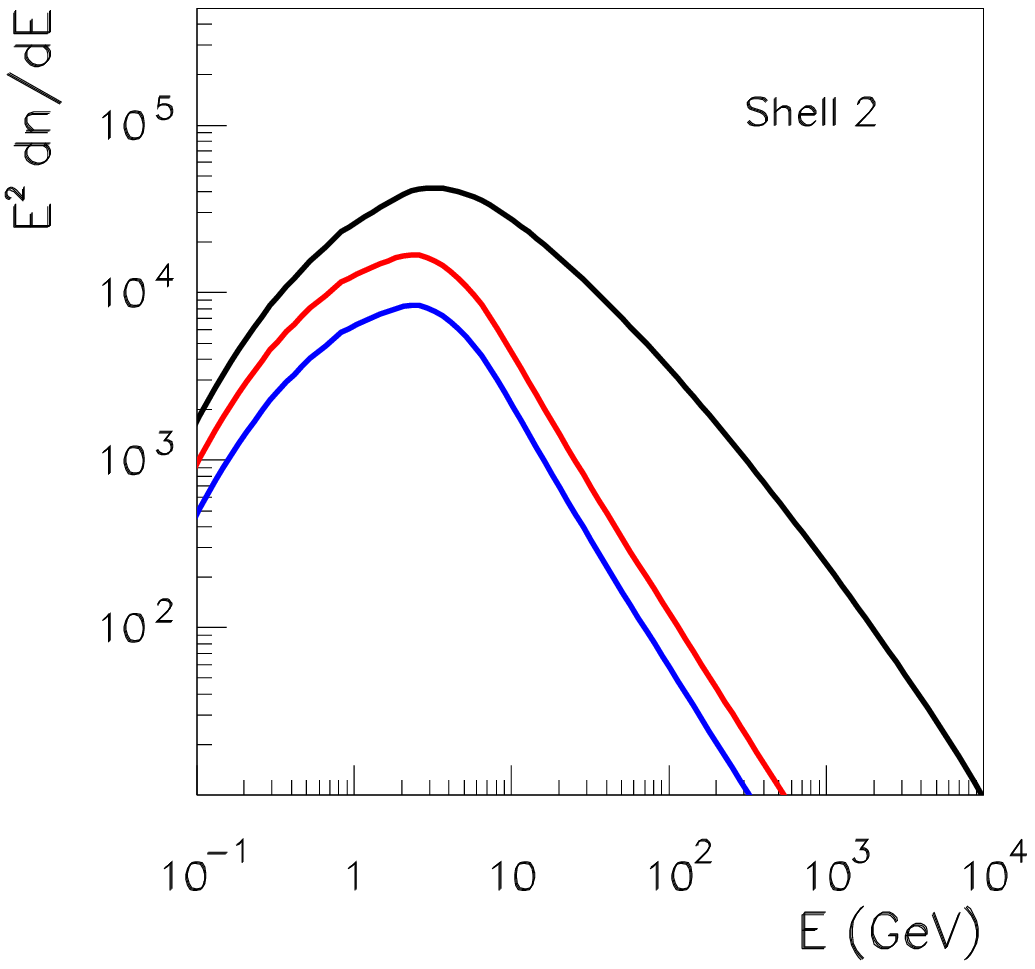}}
\vspace*{-0.4cm}
\caption{Continuous $\gamma$-ray energy spectra produced for a power-law cosmic-ray spectrum injected by a source at the origin and at $t=0$. Figure shows results for 
shells 1 \& 2 as defined in Fig.~\ref{Fig:DiffusePowerLaw}.}\label{Fig:GammaCR}
\end{center}
\end{figure*}

Finally we consider a power-law source spectrum $Q=Q_0E^{-\alpha}$ for cosmic-ray ions.
To simulate with the same statistical accuracy at all energies, each simulated particle carries a weight factor which is calculated for each energy
bin:
\begin{eqnarray}\label{EQ:WeightFn}
w(E) =Q_0E^{-\alpha}\frac{N_{\textrm{Bin}}}{N_0}\delta E
\end{eqnarray}
and the normalization factor $Q_0$ is obtained by
\begin{eqnarray}\label{EQ:WeightFn}
\int Q dE = \int Q_0 E^{-\alpha} dE = N_0
\end{eqnarray}
with $N_0$ the total particle number in the simulation and $N_{\textrm{Bin}}$ the total energy bin number.
Fig.~\ref{Fig:DiffusePowerLaw} shows the simulation results (dashed lines) and analytic solutions (solid lines) of the particle distribution 
in shells 1 and 2 at different elapsed time: in the ranges of $10^2-10^4$ years 
and $10^4-10^5$ years, respectively, after been instantaneously injected by a source at the origin and with an inelastic energy loss . Note a total of $10^7$ particles 
with energies from 10 to $10^5$ GeV are simulated in this case. The inelastic energy loss in this energy range is approximated 
by $\dot{E}\sim 5\cdot 10^{-16}E^{1.07}$ \cite{Huang07} to facilitate the analytical solution and replace the
synchrotron energy loss rate in Eqs.~(\ref{EQ:SolESynchrotron}) and (\ref{EQ:CoolingTime}). Also note the time is 
relatively short when compared with the particle propagation time ($10^8$ years), therefore we describe the particle distribution in the surroundings of sources.
\subsection{Diffuse $\gamma$-Ray Production}
By using a $\gamma$-ray production matrix published earlier \cite{Huang07}, the diffuse $\gamma$-ray spectrum produced by the 
cosmic-rays can be determined. Fig.~\ref{Fig:GammaCR} shows the continuous $\gamma$-ray spectra produced by cosmic-rays with distributions at different elapsed timescales 
shown in Fig.~\ref{Fig:DiffusePowerLaw}. Figs.~\ref{Fig:DiffusePowerLaw} and \ref{Fig:GammaCR} present an example of spatial distribution and time evolution of 
cosmic-rays and $\gamma$-rays in the source vicinity. The cosmic-ray population in the Galaxy is thought to be approximately stationary, even though occasional
spikes in the cosmic-ray flux must be expected \cite{buesching05}. This study of cosmic-ray propagation may provide a more detailed view of spectral variations in 
the vicinity of cosmic-ray sources. 

\subsection{Conclusions}
Cosmic-ray propagation in Galaxy and $\gamma$-ray production during the cosmic-ray propagation are studied by a fully 3-dimensional Monte Carlo simulation
approach. Our results are in good agreement with analytic solutions to the diffusion equation. The spatial resolution in this simulation varies between 
about 0.5 pc and about 10 pc, depending on particle energy as well as the elapsed time. The calculations may prove useful in determining the distribution 
and spectra of cosmic rays and their time evolution in the vicinity of their sources.
\subsection{Acknowledgments}
The author C.-Y. Huang gratefully thanks Ming Li for helpful discussions on the parallel computation. Grant support from NASA with award No. NAG5-13559 is 
gratefully acknowledged.

%This is the reference to .bib file (Whitout .bib!)
\bibliography{libros}

%This in the bibtex style, is ok.
\bibliographystyle{plain}
\end{document}